\documentstyle[sprocl,epsf]{article}

\newcommand{\be}{\begin{equation}}
\newcommand{\ee}{\end{equation}}
\newcommand{\beq}{\begin{eqnarray}}
\newcommand{\eeq}{\end{eqnarray}}

\begin{document}

%\vspace*{-2cm}

%\begin{flushright}
% UCY-PHY-98/04 
%\end{flushright}

%\vspace*{1.cm}

\title{Thermodynamics of One-flavour QCD\footnote[1]
{Talk presented by C. Alexandrou.} }

\author{C. Alexandrou$^a$, A. Bori\c{c}i$^b$, A. Feo$^a$, Ph. de Forcrand$^c$,
 A. Galli$^d$,
F. Jegerlehner$^e$ and T. Takaishi$^f$}
\address{$^a$Department of Natural Sciences, 
        University of Cyprus, CY-1678 Nicosia, Cyprus\\
$^b$ Paul Scherrer Institute, CH - 5232 Villigen, Switzerland\\
$^c$ SCSC,  ETH-Zentrum, CH-8092 Z\"urich, Switzerland\\
$^d$ELCA Informatique, HofwiesenStr. 26,          
       CH-8057 Z\"urich, Switzerland\\
$^e$DESY-IfH Zeuthen, D-15738 Zeuthen, Germany\\
$^f$ Hiroshima University of Economics, Hiroshima, Japan 731-01
}

\maketitle\abstracts{
We give a brief introduction 
on finite temperature phase transitions in lattice QCD including a discussion
on the identification of first order transitions. 
We present a study of the 
deconfinement phase transition  of one-flavour QCD, 
using the multiboson algorithm on lattice of sizes 
$8^3$, $12^3$ and $16^3\times 4$. 
For heavy quarks our results 
are characteristic of  a first order phase
transition which gets  weaker as the quark mass decreases 
and ends at a critical value of $\kappa \sim 0.1$ or in physical units at
about 1.6 GeV.
}

\section{Introduction}
Twenty years ago  Polyakov \cite{Polyakov} and 
Susskind \cite{Suss} pointed out that, when the temperature is increased,
  a deconfinement phase transition from 
normal hadronic matter to quark-gluon plasma can occur. 
Understanding the properties of QCD under extreme
conditions of high temperature and/or pressure has been 
ever since a challenging problem (for a recent review see for example 
Ref.~\cite{Ortmanns}).

The nature of the deconfinement phase transition has far-reaching
phenomenological consequences: In Astrophysics it is important because 
such a phase transition is believed to have occurred,
 in the opposite direction, $10^{-6} {s}$ after the Big Bang. If  of
first order, effects due to  supercooling should be visible today, possibly 
as deviations in the light element abundance from values obtained in the
 standard scenario. Determination of the order  is also of importance to 
model  builders of neutron stars and supernovae.
In experiments with ultra-relativistic heavy-ions the aim is to create and 
detect the quark - gluon plasma. A first order phase 
transition is generally considered easier to detect and many of the proposed signatures assume
the existence of a mixed phase~\cite{Greiner}.
In the planned heavy ion experiments that will start next year at RHIC, Brookhaven
and later on at LHC, CERN,
 the temperature reached will be of the order of 600 MeV 
and at this temperature one is still dealing with a strongly 
interactive system.
Thus lattice QCD provides the most suitable non-perturbative approach to study such
phenomena starting directly from the QCD Lagrangian. In order to use lattice
QCD to study these phenomena 
we need an efficient method for simulating an odd number of flavours, namely
the two light $u$ and $d$ quarks and the heavier strange quark. The local
bosonic algorithm originally proposed by L\"uscher for two degenerate 
flavours~\cite{Luscher} can be generalized to any number of 
flavours~\cite{BF,BFG}.  Before treating the $(2+1)$-flavour case, we have
applied it first to  the study of one-flavour QCD~\cite{lat,paper}.
In part because of algorithmic
difficulties,  one-flavour QCD has been largely ignored 
in spite of its interesting properties.
 The continuum $N_f=1$ theory contains no pions and is  expected to have 
no chiral phase transition\cite{PW}, whereas
the heavy quark regime of the theory
is expected to be qualitatively
similar  to  the three dimensional three-states Potts model \cite{DeGDeT}
in an external
magnetic field. 
%where  the first order 
%phase transition weakens with the external field turning  into a crossover.

In this talk, after giving a brief overview of how one studies first-order
phase transitions on a finite lattice taking as an example the well-studied
deconfinement phase transition in quenched QCD, we will present
results for heavy to moderately heavy Wilson fermions for one flavour QCD,
on lattices of size  
$8^3$, $12^3$ and $16^3 \times 4$.

\section{Overview}
  The deconfinement phase transition was first studied in the case 
of pure gauge. The basic degrees of freedom here are the $SU(3)$ link variables
$U_{\mu}(n)$ located on the link leaving
site $n$ in the direction $\hat{\mu}$ 
of a space-time lattice with $N_s^3$ spatial 
size and temporal size $N_t$, with $N_s \gg N_t$ if possible. The temperature is given
by $T=(N_t a)^{-1}$ with $a$ the lattice spacing.
The partition
function in Euclidean space is given by 
\be
Z= \int \>\>[dU] \>\> e^{-S_G} \hspace*{0.5cm} {\rm and} \hspace*{0.5cm}
S_G=\beta \sum_P\>[1-\frac{1}{3}\>{\rm Re \> Tr}\>U_P]
\label{pure gauge Z}
\ee
where $U_P$ is the product 
$U_{\mu}(n)U_{\nu}(n+\hat{\mu})U_{\mu}^{\dagger}(n+\hat{\nu})U_{\nu}^{\dagger}(n)$
of $4$ links around an elementary plaquette $P$,
and we sum over all plaquettes $P$. 
We take periodic  boundary conditions for the links and,
for a given $N_t$, the temperature is changed by varying $\beta=6/g^2$
with $g$ the coupling constant.
Svetitsky and Yaffe~\cite{YS} identified the global
symmetry that is spontaneously broken at the deconfinement phase transition
as the center symmetry $Z(3)$ of $SU(3)$. The pure gauge
action is invariant under $Z(3)$. To see this let us transform 
all time-like links with a given time coordinate $n_0$,
\be
U_0({\bf n},n_0) \rightarrow z U_0({\bf n},n_0)
\label{Z(3) transformation}
\ee
 with $z$   %$z=\exp(2i\pi m/3), m=1,2,3$
 an element
of $Z(3)$. A temporal plaquette at time $n_0$  
necessarily has two
transformed links so that $U_P\rightarrow U_{\hat{\mu}}(n) z U_0(n+\hat{\mu})
U_{\mu}^{\dagger}(n+\hat{0}) U_0^{\dagger}(n)z^{\dagger}=U_P$ since
$z$ commutes with all $U$'s. While local observables are invariant, the Polyakov
line defined as the product of temporal links
\be
\label{polyakov_loop}
L({\bf n}) = \frac{1}{3} {\rm Tr} \prod_{n_0=1}^{N_t} U_0({\bf n},n_0) \quad,
\ee
transforms as $L \rightarrow zL$ under $Z(3)$. Its
 expectation value $\langle L\rangle=0$ if the symmetry is unbroken whereas 
$\langle L\rangle \neq 0$
if the symmetry is spontaneously broken. If $\langle L\rangle=0$
 the free energy of a static 
quark, $F_q=-\log(\langle L\rangle)/\beta $, becomes infinite   and  
at large distances the correlator 
$G({\bf r}) \equiv  \langle L({\bf n})L({\bf n}+{\bf r})\rangle
\sim \exp(-\beta F_{q\bar{q}})$ 
should also vanish with an exponential dependence
$\sim \exp(-\sigma a^2 |{\bf r}|N_t)$
with $\sigma$ the string tension and  $F_{q\bar{q}}$ the 
free energy of a static quark-antiquark pair~\cite{Sve}.
 This behaviour describes
the confining phase. If the symmetry is spontaneously broken then $\langle L\rangle \neq 0$
and $F_q$ is finite signaling quark liberation. Therefore the Polyakov loop is
an order parameter for the $Z(3)$ symmetry. Constructing an effective action 
for $L$ and using universality arguments the authors of  Ref.~\cite{Sve}
 predicted that the phase transition in pure gauge
is first order. This has been confirmed by numerical lattice 
simulations~\cite{Fuk}.

Dynamical quarks are included using the standard Wilson discretization 
procedure. Writing the fermion matrix $D=1-\kappa M$, where $M$
is the hopping matrix, the Wilson action is

\vspace*{-0.8cm}

\beq
\nonumber
S_F^{(W)} &=&\sum_{nn'} \bar{\psi}(n) D(n,n')\psi(n'); \hspace*{0.4cm}
 2m_q a=\frac{1}{\kappa}-\frac{1}{\kappa_c}; \hspace*{0.4cm}
M \equiv \sum_{\mu=0}^3 M_\mu; 
\hspace*{0.4cm}
\\
M_{\mu}(n,n')&=&(1-\gamma_{\mu})U_{\mu}(n)\delta_{n',n+{\hat{\mu}}} 
       + (1+\gamma_{\mu})U_{\mu}^{\dagger}(n') \delta_{n',n-{\hat{\mu}}} \quad.
\label{Wilson action}
\eeq 
The hopping parameter $\kappa$ is thus related to the quark mass, $m_q$,
and $\kappa_c$ is the value of $\kappa$ for the massless limit.
For the fermions we use antiperiodic boundary conditions in the temporal 
direction and periodic in the spatial directions. 
Under the $Z(3)$ transformation (\ref{Z(3) transformation}),
$\bar{\psi}({\bf n},n_0)U_0({\bf n},n_0)\psi({\bf n},n_0+\hat{0})
\rightarrow z\bar{\psi}({\bf n},n_0)U_0({\bf n},n_0)\psi({\bf n},n_0+\hat{0})$ 
and $S_F^{(W)}$ is no longer invariant. 
This explicit symmetry breaking
can weaken the first order phase transition, change it to second order or even make it
disappear. These possibilities must be investigated by detailed calculations
as a function of the quark mass. The order parameter $\langle L \rangle$ is no longer zero
even in the ``confined'' phase since we can have  $q\bar{q}$ pair creation shielding
colour. However it will still show a discontinuity at the phase transition
and can still be used as  a probe for deconfinement.
In the partition function the Wilson fermions can be integrated out yielding
\be
Z= \int\> d[U] \exp(-S_G^{\rm eff}[U]) \hspace*{0.5cm} {\rm with} 
 \hspace*{0.5cm} S_G^{\rm eff}=S_G-{\rm Tr log}(1-\kappa M) \quad.
%\int \>d{U}\>d\bar{\psi}\>d[\psi] \>\exp(-S_G[U]-S_F^{(W)}) 
\label{Z with fermions}
\ee
$S_G^{\rm eff}$ can be 
expanded in powers of $\kappa$ and for finite temperature the $\kappa^{N_t}$
term gives a contribution to $S_G^{\rm eff}$ of the form~\cite{Sve}
\be
S_h=-h\>\sum_{\bf n}\>\>{\rm Tr\> Re}\> L({\bf n}), \hspace*{0.4cm} h > 0
\label{S_h}
\ee 

The effective magnetic field, $h \propto \kappa^{N_t}$ to leading order,
 induces
a magnetization along the positive real axis. Since the transition at $h=0$ is
first order the magnetic term is expected to weaken it and above a critical
value $h_{ep}$ the transition may disappear.

\begin{figure}[h]
\begin{minipage}{6.5cm}
\parbox{6.5cm}{ 
%The typical phase diagram in this
%case is shown in fig.~\ref{fig:bc_diag} and has been studied 
Such a behaviour was observed in the three 
dimensional three-state Potts model in an external magnetic field~\cite{DeGDeT} $h$:
\be 
S= - \sum_{<nl>} \>\beta \>{\rm Re}\> z^*_n z_l - h\> \sum_n\> {\rm Re}\> z_n
\label{S Potts}
\ee
with $z_n$ an element of $Z(3)$.
Whereas Fig.~\ref{fig:bc_diag} may indicate
 the expected qualitative behaviour for QCD with a dynamical quark, 
the question of existence of an end-point
% $h_{ep}$
 and its 
location can only be answered after a detailed
calculation.
}
\end{minipage} \hfill
\begin{minipage}{4.5cm}
\mbox{\epsfxsize=5cm\epsfysize=4cm\epsffile{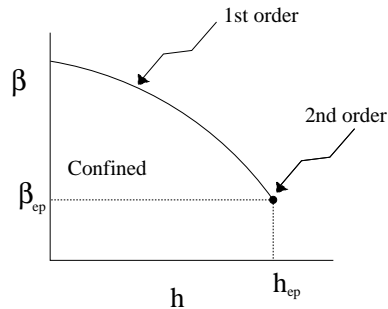}}
\vspace*{-0.7cm}
\caption{Phase diagram found in the three
 dimensional three-states Potts model in a magnetic field $h$}  
\label{fig:bc_diag}
\end{minipage}
\end{figure}

\vspace*{-0.5cm}

\section{Local bosonic algorithm for one flavour}
The generalization of the local bosonic algorithm~\cite{Luscher}
to any number of flavours is made easier by finding
a polynomial approximation to the fermionic matrix itself~\cite{BF,BFG}
 rather than to
$(\gamma_5 D)^2$.  
This can be done by constructing a polynomial 
of even degree $n$  
defined in the complex plane with complex conjugate roots $z_k$ 
such that
%\be
$\lim_{n\rightarrow \infty} P_n(z)= \frac{1}{z}$
%\label{P_n complex}
%\ee
for any $z$ in the  domain containing the spectrum of $D$ 
(not including the origin). 
Since the spectral radius of the hopping matrix $M$ is bounded by 8
in the free case, less in the interacting one, we are guaranteed that
the spectrum of $D = 1 - \kappa M$ will remain in the complex right half-plane
for the heavy to moderately heavy quarks we simulate ($\kappa \leq 1/8$).
Using the property $D=\gamma_5 D^{\dagger} \gamma_5$ we can write
\be
det P_n(D)=c_n \prod_{k=1}^{n/2} det(D-z_k)^{\dagger}det(D-z_k)
\label{det P_n }
\ee
with $c_n$ an easily computed  constant,\cite{BF}
 obtaining a local representation for $det D$
\be
det D = \lim_{n\rightarrow \infty} det^{-1} P_n(D) 
        =  \int\prod_{k=1}^{n/2} d\phi^{\dagger}_k d\phi_k 
         e^{-\sum_{k=1}^{n/2} \phi^{\dagger}_k 
                          (D-z_k)^{\dagger}(D-z_k) \phi_k} \quad 
\label{det D}
\ee
The algorithm is made exact with a global Metropolis test~\cite{BFG,paper}
at the end of a ``trajectory''.
The number $n$ of bosonic fields $\phi_k$ is chosen so that the correction term
leads to an acceptance  rate of about 2/3.

For the local updating of the gauge and boson fields we used standard heatbath
and over-relaxation algorithms as in Ref.~\cite{BFG}.
A trajectory 
is a symmetric combination of $(2 m + 1)$ over-relaxation steps applied 
alternatively to the gauge and boson fields, preceded and followed by
a heatbath on the bosons.
%here consists of a heatbath for the bosons,  two sequences of
%over-relaxation of the bosons and gauge fields, symmetrized with an additional
%over-relaxation and heatbath of the bosons. 
Ergodicity for the gauge fields is
maintained due to their coupling to the bosonic fields. 
 The roots $z_k$ are
distributed on the circle of radius 1 centered at (1,0). 
We use even-odd preconditioning 
 to lower the number of bosonic fields needed for a given accuracy, and 
 a quasi-heatbath
 to initialize
the boson fields using thermalized gauge configurations from other
$\kappa$ and $\beta$ values \cite{paper}.

\section{Order parameters}
In section (2) we argued that the deconfinement
phase transition is expected to be first order for heavy quarks.    
 Since the discontinuities which characterize a first
order phase transition in the continuum are smoothed out on a finite lattice,
we must rely on finite size scaling to identify a first order transition.
Finite size scaling was first applied to identify transitions 
 in  spin systems 
\cite{ising} and later 
to identify the order of the deconfinement phase transition 
in quenched QCD~\cite{Fuk}.
The observables that we consider are based on the Polyakov 
loop~(\ref{polyakov_loop}).

Our strategy is to vary $\beta$ at  a given quark mass for our three spatial
lattice sizes, and
 look for the following signals:\\ 
%\begin{itemize}
%\item {\it Coexistence of the two phases:}
{\it Coexistence of the two phases:}
A distinctive feature of a first order transition is phase coexistence.
On a finite lattice  
 we can look for tunneling between 
confined and deconfined phase, which occurs over a small temperature
range around the  critical temperature. 
%As the size of the lattice increases
%tunneling is suppressed but for the lattice sizes studied here enough 
We observe enough tunneling events to study the
  double peak distribution of the norm $|\Omega|$ of the Polyakov loop
 defined as 
\be
\Omega=\frac{1}{V} \sum_{\bf n}L(\bf{n}) 
\label{Omega}
\ee
with $V$ the spatial volume.
%The width of the peaks should scale like $V^{-1}$ for a first order phase 
%transition just as in the case of the 2-d Ising model \cite{ising}.

\noindent
%\item {\it Deconfinement ratio:} The deconfinement ratio is defined by 
{\it Deconfinement ratio:} The deconfinement ratio is defined by 
\be
\rho = \frac{3}{2}\> p -\frac{1}{2}
\label{rho}
\ee
 with $p$ the probability for the complex Polyakov loop to be within $20$ degrees
of a $Z_3$ axis. Therefore if the $Z(3)$ symmetry is unbroken we find  $p=1/3$
and $\rho=0$, while if it is broken in such a way that the Polyakov loop is 
distributed around one axis (here the real axis), then 
$p=1$ and $\rho=1$.   The value $\rho=0$ 
is obtained only in the pure gauge case
 where the  $Z(3)$ symmetry
 is exact. Since heavy
quarks break $Z(3)$  we obtain a value of $\rho$ different 
from zero in the confined phase but still look for a discontinuity across the
phase transition.
% but still a remnant $Z(3)$ symmetry
% leads to a
%$\rho$ value smaller than unity.

\noindent
%\item {\it The peak value of the susceptibility:}  The susceptibility gives
{\it The peak value of the susceptibility:}  The susceptibility gives
a measure of the fluctuations of the Polyakov loop.
We consider the behaviour of
\be
\chi_L=V(\langle|\Omega|^2\rangle-\langle|\Omega|\rangle^2)
\label{chi}
\ee
which diverges 
 at criticality for a first order phase transition
in the continuum.
On a finite lattice the discontinuity is smoothed:
the distribution of $\chi_L$ has a width proportional to $1/V$
and a peak value $\chi_L^{peak} \propto V$.
This scaling behaviour  changes for a
 second order transition to  $\chi_L^{peak} \propto V^{\alpha}$ 
with $\alpha<1$. 
For a crossover behaviour, where
no discontinuity occurs even in the thermodynamic limit $V\rightarrow\infty$,
$\chi_L^{peak}$ remains constant as $V$ increases.
%\end{itemize} 

\section{Results}
 The parameters of our simulations
are included in Table~1. 
From the
values listed in the table, we find that
$n \propto log(volume)$  for a fixed quark mass and
approximately $ n \propto 1/m_q$ for a fixed volume, as expected \cite{BFG}.

%\newpage

\begin{figure}
%\vspace*{-1cm}
\begin{minipage}[t]{12cm}
\begin{minipage}[t]{5.5cm}
\parbox{5.5cm}{Our results for the observables which probe the order of
the transition are shown in 
Figs.~(\ref{fig:tunneling_k010}-\ref{fig:chi_peak}). 
Tunneling  between 
the confined and the deconfined phase is clearly observed 
%for $\kappa=0.05$ as well as 
for $\kappa=0.1$ as shown in Fig.~\ref{fig:tunneling_k010} . 
Similar results are obtained for $\kappa=0.05$ whereas for
 $\kappa=0.14$ tunneling is no longer observed. 
In Fig.~\ref{fig:r_deconf} we show the deconfinement 
ratio $\rho$ obtained using
reweighting~\cite{FeSwend}
for $\kappa=0.05, \kappa=0.10$ and 
$\kappa=0.12$. We include for comparison the pure gauge results as well.
Across the transition region the slope of $\rho$
increases with the volume for $\kappa=0.05$ signaling a first order transition
in  contrast
\vspace*{0.2cm}}
%\\}
\end{minipage}\hfill
\begin{minipage}{6cm}
\mbox{\epsfxsize=6cm\epsfysize=6cm\epsffile{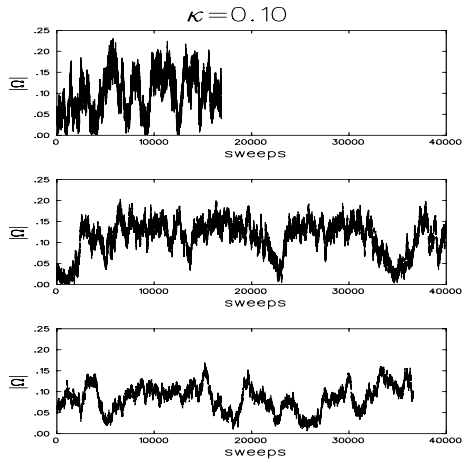}}
\caption{ Time history of $|\Omega|$ for $\kappa=0.10$ for $N_s=8$ (upper)
$N_s=12$ (middle) and $N_s=16$ (lower)}
\label{fig:tunneling_k010}
\end{minipage}
\end{minipage}
\begin{minipage}{12cm}
\parbox{12cm}{  to
 the behaviour of $\rho$ for
$\kappa=0.12$ indicating that $\kappa=0.12$ is
already in the crossover region. For $\kappa=0.10$ 
it is not very clear and one would  need a bigger lattice
comparable to the correlation length  
to make a definite decision.
The volume dependence of the 
peak value of $\chi_L$ is  displayed 
in Fig.~\ref{fig:chi_peak}.  
The lines shown are best fits to the form $V^{\alpha}$. For $\kappa=0.05$ 
the best fit yields $\alpha=0.96(4)$ whereas for $\kappa=0.14$ 
$\alpha=0$ and thus we have further evidence that for $\kappa=0.05$ the
transition is first order whereas for $\kappa=0.14$ we have a crossover. 
For $\kappa=0.10$ and $\kappa=0.12$ the situation is
less clear. The small value of $\alpha=0.22(3)$ at $\kappa=0.12$
as well as the absence of tunneling leads us to conclude that $\kappa=0.12$
is in the
crossover region. 
For $\kappa=0.10$ tunneling is still observed
but $\alpha=0.56(3)$
so that we may conclude that we are near the end point of the first order
phase transition.

\hspace*{0.5cm}The critical values $\beta(\kappa)$, given in Table~1
%\ref{table:results}
%and plotted in fig. \ref{fig:results} 
are obtained from the position of
the peak value of $\chi_L$. For $\kappa=0.05$ and $\kappa=0.1$
the value of $\beta$ where the deconfinement ratios for various volumes cross
yields a value consistent with the one obtained from the position
of $\chi_L^{peak}$. 
Not surprisingly,
the shift $\beta_c(\kappa) - \beta_c(\kappa=0)$ from the pure gauge critical 
coupling is approximately half that of the two-flavour case.
Taking the end-point value of $\kappa$, $\kappa_{ep} \sim 0.1$ with
20\% uncertainty, we can approximately
map to physical units: Using the tadpole-improvement property 
$\kappa_c(\beta) \langle plaq \rangle^{1/4} \approx 1/8$ we obtain 
$m_q a \sim 1.8$, with 
} 
\end{minipage}
\end{figure}

\noindent
$(4 a)^{-1} \sim 220 MeV$ from the deconfinement 
temperature. This gives $m_q \sim 1.6 GeV$ at the end-point.

\vspace*{0.4cm}
\begin{table}[h]
\begin{center}
\begin{tabular}{|c||cc|cc|cc||c|}
\hline
$\kappa$ &\multicolumn{2}{c}{ $8^3 \times 4$} &
    \multicolumn{2}{c}{$ 12^3 \times 4$} & 
\multicolumn{2}{c||}{$ 16^3 \times 4$} & $\beta_c$\\
\hline
\hline
 & $n/acc$ & $Ksw$ & $n/acc$ & $Ksw$ & $n/acc$ & $Ksw$ &\\
\hline
 0.05 & 8/0.78 & 18 & 12/0.74 &20 & 24/0.83 & 20 & 5.690(2)\\
\hline
0.10 & 16/0.67 & 45 & 24/0.63 & 50 & 32/0.67 & 37 & 5.66(1)\\
\hline
0.12 & 24/0.74 &55 & 32/0.67 & 40& 40/0.69 & 12 & 5.63(1)\\
\hline
0.14 & 32/0.77 & 60 & 40/0.70 &37 & 50/0.67 & 12 & 5.59(1)\\
\hline
\end{tabular}
\end{center}
\vspace*{0.4cm}
\small
Table 1: We give the $\kappa$ values, the number of bosonic
fields $n$ and the average acceptance $acc$ for the three 
volumes studied. $Ksw$  (in kilo sweeps) is 
the total number 
of thermalized configurations used in the
reweighting procedure \cite{FeSwend}. In the last column  the
critical $\beta$ values obtained from the analysis 
of the results are listed.\\
\normalsize
\end{table}
\vspace*{-0.6cm}

\section{Summary and Conclusions}
In this work we have shown that the multiboson algorithm is well-suited
for the study of one-flavour QCD for moderately 
heavy Wilson quarks. 
%For global observables like the polyakov loop
%it performs about 10 times better as compared to HMC. 
Using an exact algorithm for one flavour we have carried out
a detailed  finite size scaling analysis and determined the critical 
value of $\beta$ for $\kappa$ up to $0.14$. 

We have presented conclusive evidence that the deconfinement phase transition
is first order for heavy quarks, with the critical line ending at
 about $\kappa=0.1$, i.e. $m_q \sim 1.6$~GeV.
This is in line with phenomenological expectations. 
The pure gauge deconfinement
transition is fairly weak, with a correlation length 
$\cal O$(a few $\sigma^{-1/2}$). This is the minimum system size necessary
to observe the deconfinement transition.
Dynamical quarks introduce a new length
scale, the distance where the string breaks, 
${\cal O}(2 m_q /\sigma)$. Confinement can only be observed up to
this distance. 
When the quark mass is lowered sufficiently that the second length-scale 
is similar to (or smaller than) the first, one cannot tell if the system
is confined or deconfined, and the transition is replaced by a crossover.
This occurs for
\be
m_q \sim {\cal O}(a few \sqrt\sigma / 2) 
\hspace*{1cm} i.e.\hspace*{1cm}  m_q \sim {\cal O}(1) GeV
\label{end point mass}
\ee

{\bf Acknowledgements:} We thank SIC of the EPFL in
Lausanne, ZIB in Berlin, and  the Minnesota Supercomputing Institute
for computer time.

\newpage

\begin{figure}[t]
\begin{minipage}{5.5cm}
\mbox{\epsfxsize=6cm\epsfysize=6cm\epsffile{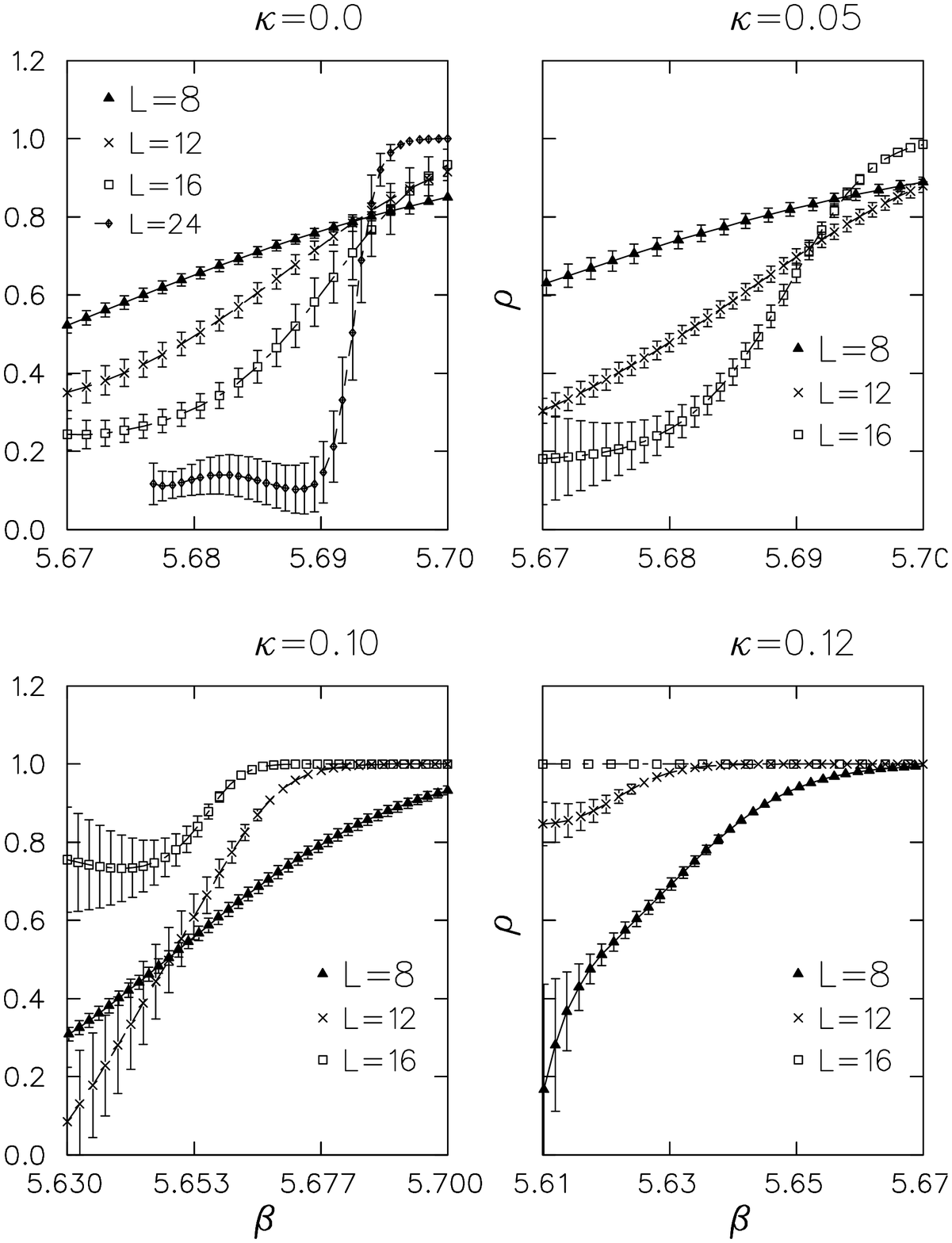}}
%\end{center}
\caption{The Deconfinement ratio for pure gauge and for $\kappa=0.05$, $\kappa=0.10$ and $\kappa=0.12$ for three lattice sizes}
\label{fig:r_deconf}
\end{minipage} \hfill
\begin{minipage}{5.5cm}
\mbox{\epsfxsize=6cm\epsfysize=6cm\epsffile{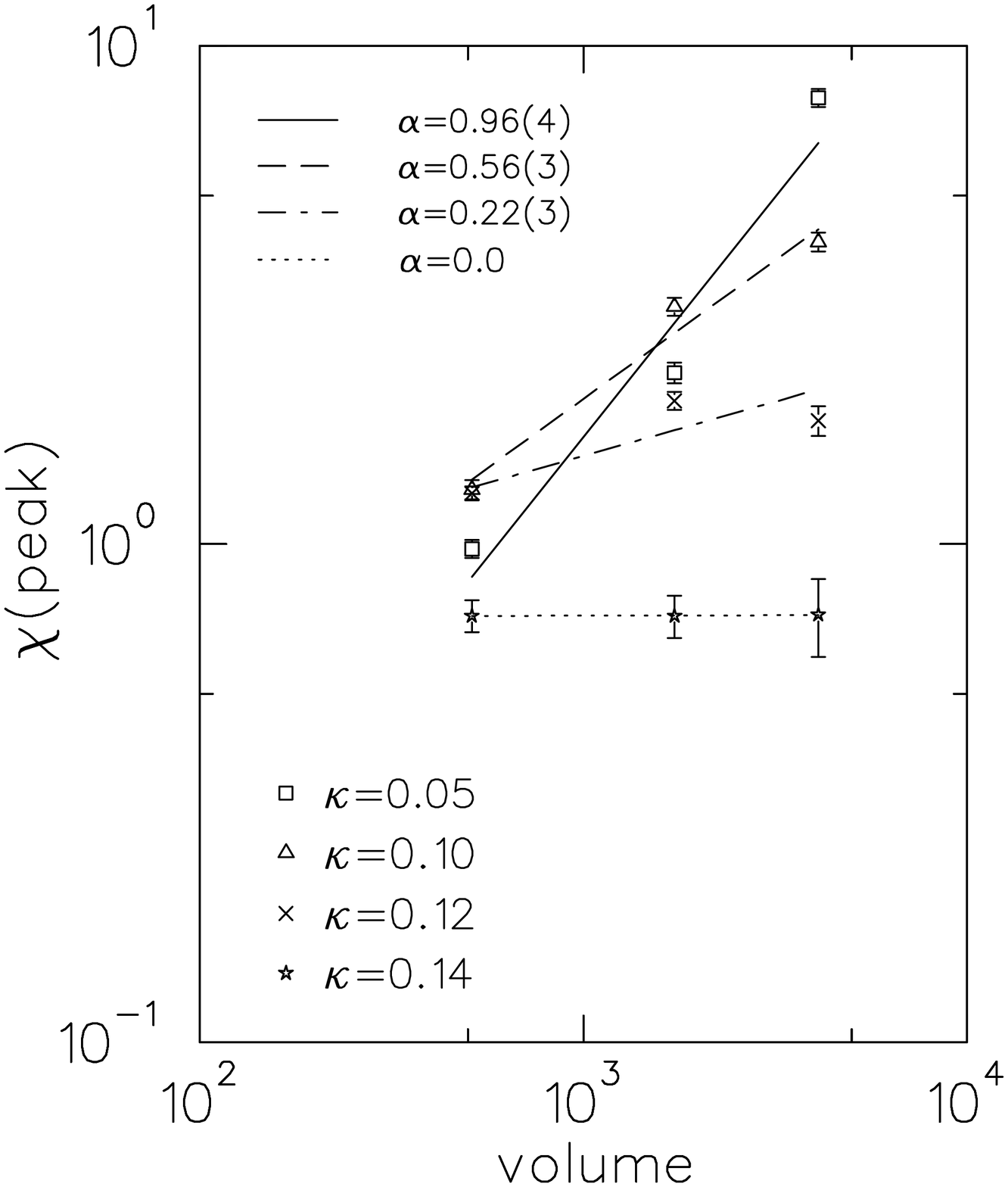}}
\caption{The volume dependence of the peak of the susceptibility for $\kappa=0.05$, $\kappa=0.10$, $\kappa=0.12$ and $\kappa=0.14$.}
\label{fig:chi_peak}
\end{minipage} 
\end{figure}

\end{document}